\documentclass[showpacs,prl,floatfix,preprintnumbers,twocolumn]{revtex4}
\usepackage{graphicx,url,amssymb,amsmath,longtable,rotating,color,psfig,units,wasysym,subfigure,epsfig,listings,bm}

\begin{document}
\vspace{1cm}

\title{Parameter Estimation in Searches for the Stochastic Gravitational-Wave Background}

\author{V. Mandic$^a$, E. Thrane$^{a}$\footnote{ethrane@physics.umn.edu}, S. Giampanis$^b$ and T. Regimbau$^c$}
\affiliation{$^a$School of Physics and Astronomy, University of Minnesota, Minneapolis, MN 55455, USA\\
$^b$University of Wisconsin-Milwaukee, Milwaukee, WI  53201, USA\\
$^c$Departement Artemis, Observatoire de la C\^ote d'Azur, CNRS, F-06304 Nice,  France}

\date{\today}

\begin{abstract}
  The stochastic gravitational-wave background (SGWB) is expected to arise from the superposition of many independent and unresolved gravitational-wave signals of either cosmological or astrophysical origin. The spectral content of the SGWB carries signatures of the physics
  that generated it. We present a Bayesian framework for estimating the parameters associated with different SGWB models using data from gravitational-wave detectors. We apply this technique to recent results from LIGO to produce the first simultaneous $95\%$ confidence level limits on multiple parameters in generic power-law SGWB models and in SGWB models of compact binary coalescences.
  We also estimate the sensitivity of the upcoming second-generation detectors such as Advanced LIGO/Virgo to these models and demonstrate how SGWB measurements can be combined and compared with observations of individual compact binary coalescences in order to build confidence in the origin of an observed SGWB signal. In doing so, we demonstrate a novel means of differentiating between different sources of the SGWB.
\end{abstract}


\bibliographystyle{unsrt_modified_2}
\maketitle

{\em Introduction.}---The stochastic gravitational-wave background (SGWB) is expected to arise from the superposition of gravitational waves (GWs) from many uncorrelated and unresolved sources.
Numerous cosmological and astrophysical models have been proposed.
Cosmological models include the slow-roll inflation model~\cite{grishchuk,starob}, a variety of inflationary models with significant boosts in energy density at high frequencies (parametric resonance in the preheating phase \cite{eastherlim}, models of axionic natural inflation \cite{peloso}), models based on cusps or kinks in cosmic (super)strings \cite{caldwellallen,DV1,DV2,cosmstrpaper,olmez1,olmez2}, and models of alternative cosmologies such as pre-Big-Bang models \cite{PBB1,PBBpaper}.
Astrophysical models integrate contributions from various astrophysical objects across the universe including compact binary coalescences (CBC) of  binary neutron stars (BNS) or binary black holes (BBH) \cite{phinney,kosenko,regimbau,zhu,StochCBC}, magnetars \cite{cutler,marassi,RegMan}, rotating neutron stars \cite{RegPac,owen,barmodes1,barmodes2,barmodes3}, the first stars \cite{firststars}, and white dwarf binaries~\cite{phinney_whitedwarfs}.

In all models, the SGWB is described in terms of the normalized GW energy density,
\begin{equation}
  \Omega_{\rm GW}(f) = \frac{f}{\rho_c} \; \frac{d \rho_{\rm GW}}{df}\,,
\end{equation}
where $d\rho_{\rm GW}$ is the energy density of GWs in the frequency range $f$ to $f+df$
and $\rho_c$ is the critical energy density of the universe \cite{allenromano}. The amplitude and the frequency dependence of this GW spectrum depend on the physics of the model that generated it.
For example, in the CBC model the spectrum is determined by the rate of binary systems throughout the universe and by the distribution of their chirp masses.

Past SGWB searches, performed using data from the first-generation interferometric GW detectors LIGO \cite{LIGOS1,LIGOS5} and Virgo \cite{Virgo1}, assumed a power-law model,
\begin{equation}\label{eq:powerlaw}
  \Omega_\text{GW}(f)=A(f/f_\text{ref})^\alpha ,
\end{equation}
and set limits only on the amplitude $A$ for fixed values of the spectral index $\alpha$ and of the reference frequency $f_\text{ref}$~\cite{S3stoch,S4stoch,S5stoch,s5vsr1}.
In this Letter we introduce a Bayesian technique, building on \cite{emma_bayes}, to simultaneously estimate multiple parameters of SGWB models using cross-correlation data from pairs of GW detectors~\footnote{In~\cite{emma_bayes}, the authors consider a power-law model similar to the problem considered here, as well as cases involving interferometers with correlated noise, and cases with anisotropic backgrounds. In this paper we follow a similar Bayesian formalism, but we adopt an implementation that in a simple manner extends existing stochastic data analysis pipeline, and we apply it to recent LIGO data to make first measurements of multiple parameters for specific theoretical models.}.
We apply this technique to recent results from LIGO to produce the first simultaneous limits on multiple parameters for power-law and CBC models of the SGWB. We also estimate the sensitivity of second-generation GW detectors to these models---Advanced LIGO (aLIGO) \cite{aLIGO2}, Advanced Virgo \cite{aVirgo}, GEO-HF \cite{GEOHF}, and KAGRA \cite{CLIO,LCGT} are expected to produce first results in 2014, and will be sufficiently sensitive to probe a variety of interesting SGWB models~(see, e.g.,~\cite{StochCBC}). We demonstrate that the technique can be easily extended to include other measurements, such as direct GW observations of resolvable compact binary coalescences, in order to better constrain astrophysical parameters and to gain insight into the origin of the observed SGWB signal.
%

{\em Method.}---We define the cross-correlation estimator
\begin{equation}
  \hat Y = \frac{T}{2}  \int df \tilde{s}^*_1(f)\tilde{s}_2(f) \tilde{Q}(f)\,,
  \label{ptest}
\end{equation}
where $T$ is the measurement time, $\tilde{s}_1(f)$ and $\tilde{s}_2(f)$ are Fourier
transforms of the strain time-series of two GW detectors, and $\tilde{Q}(f)$ is a filter \cite{allenromano}.
In the small-signal approximation, and assuming stationary, Gaussian noise---uncorrelated between the two detectors---the variance
of $\hat Y$ is given by:
\begin{equation}
  \sigma_Y^2 = \frac{T}{4} \int df P_1(f) P_2(f) |\tilde{Q}(f)|^2\,,
  \label{sigma}
\end{equation}
where $P_i(f)$ are the one-sided strain power spectral densities of the two GW detectors.
Optimization of the signal-to-noise ratio (SNR) leads to the following optimal filter for a frequency-independent GW spectrum $\Omega_\text{GW}(f) = \Omega_0$~\cite{allenromano}:
\begin{equation}
  \label{optfilt1}
  \tilde{Q}(f) = \mathcal{N} \; \frac{\gamma(f)}{f^3 P_1(f) P_2(f)} \; ,
\end{equation}
where $\gamma(f)$ is the overlap reduction function arising from the overlap of antenna patterns of GW detectors at different locations and with different orientations \cite{allenromano}.
The normalization constant $\mathcal{N}$ is chosen so that  $\langle \hat{Y}\rangle = \Omega_0$.

Most proposed SGWB models $\Omega_\text{M}(f|\vec\theta)$ vary slowly with frequency; (here $\vec\theta$ denotes the model parameters). It is therefore a good approximation to compute the above estimator for a series of small, $\unit[0.25]{Hz}$-wide frequency bins (see, e.g.,~\cite{S5stoch}), and define the following likelihood function:
%
%
\begin{equation}
  L(\hat Y_i, \sigma_i|\vec\theta) \propto \exp \left[ -\frac{1}{2}
    \sum_i \frac{(\hat Y_i - \Omega_{\rm M}(f_i; \vec\theta))^2}{\sigma_i^2} \right]
\label{sgwblik}
\end{equation}
where the sum runs over frequency bins $f_i$, and $\hat Y_i$ and $\sigma_i^2$ are the estimator and variance in the frequency bin $f_i$ given respectively by Eqs. \ref{ptest} and \ref{sigma}. Since $\hat Y_i$ is calculated by averaging over many short time segments, its distribution is expected to be Gaussian due to the central limit theorem, as observed, e.g., in \cite{S5stoch}.
Multiplying the likelihood with the prior distribution for $\vec\theta$ yields the Bayesian posterior distribution for the free parameters $\vec\theta$, which can then be used to extract confidence intervals for $\vec\theta$.
In the subsequent results, we take all priors to be flat within the plotted range of $\vec\theta$ (and zero elsewhere).

\begin{figure}
  \psfig{file=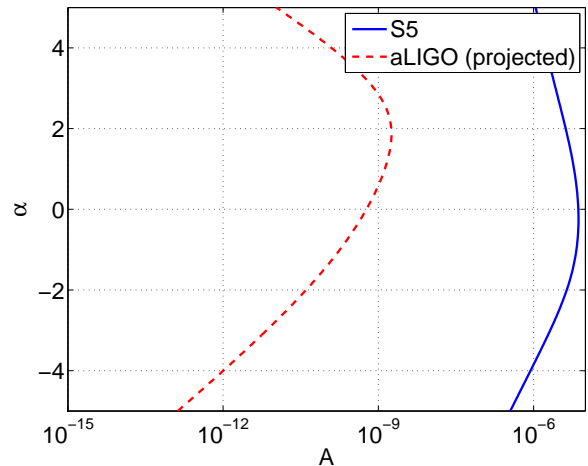, height=2.5in}
  \caption{
    The solid curve denotes the 95\% CL limit for the power-law SGWB model $\Omega_{\rm M}(f; A,\alpha)=A(f/\unit[100]{Hz})^\alpha$, in the $\alpha$-$A$ plane using the latest SGWB measurement with LIGO detectors \cite{S5stoch}. The dashed curve denotes the projected sensitivity obtainable by aLIGO at ${\rm SNR} = 2$, assuming $\unit[1]{yr}$ of data from two colocated detectors operating at design sensitivity (assuming $\hat Y_i=0$).
  }
  \label{fig:S5powerlaw}
\end{figure}

{\em Power-law spectrum.}---As most SGWB models predict power-law dependence in the LIGO frequency band, we apply the above formalism to the power-law model of the SGWB (see Eq.~\ref{eq:powerlaw}) with a reference frequency of $f_\text{ref}=\unit[100]{Hz}$.
In Fig.~\ref{fig:S5powerlaw} we use recent LIGO measurements of $\hat Y_i$ and $\sigma_i$ \cite{S5stoch} to place $95\%$ confidence level (CL) limits on $\alpha$ and $A$.
These limits are the first to simultaneously constrain two parameters of an SGWB model.
We also calculate the projected sensitivity for colocated aLIGO detectors that would be obtained after $\unit[1]{yr}$ of exposure at the design strain sensitivity and with ${\rm SNR} = 2$ (assuming $\hat Y_i=0$).

{\em Compact binary coalescences.}---Coalescences of binary systems such as BNS and BBH are among the most promising sources of GWs in the $\unit[10-1000]{Hz}$ frequency band.
The nearest CBCs are expected to produce GW signals strong enough to be individually detected by second-generation detectors~\cite{cbc_rates}.
Meanwhile, integrating contributions from all CBCs in the universe leads to an SGWB that may also be detected by cross-correlating pairs of GW detectors \cite{phinney,kosenko,regimbau,zhu,StochCBC,rosado}.

\begin{figure*}[!t]
   \begin{tabular}{cc}
    \psfig{file=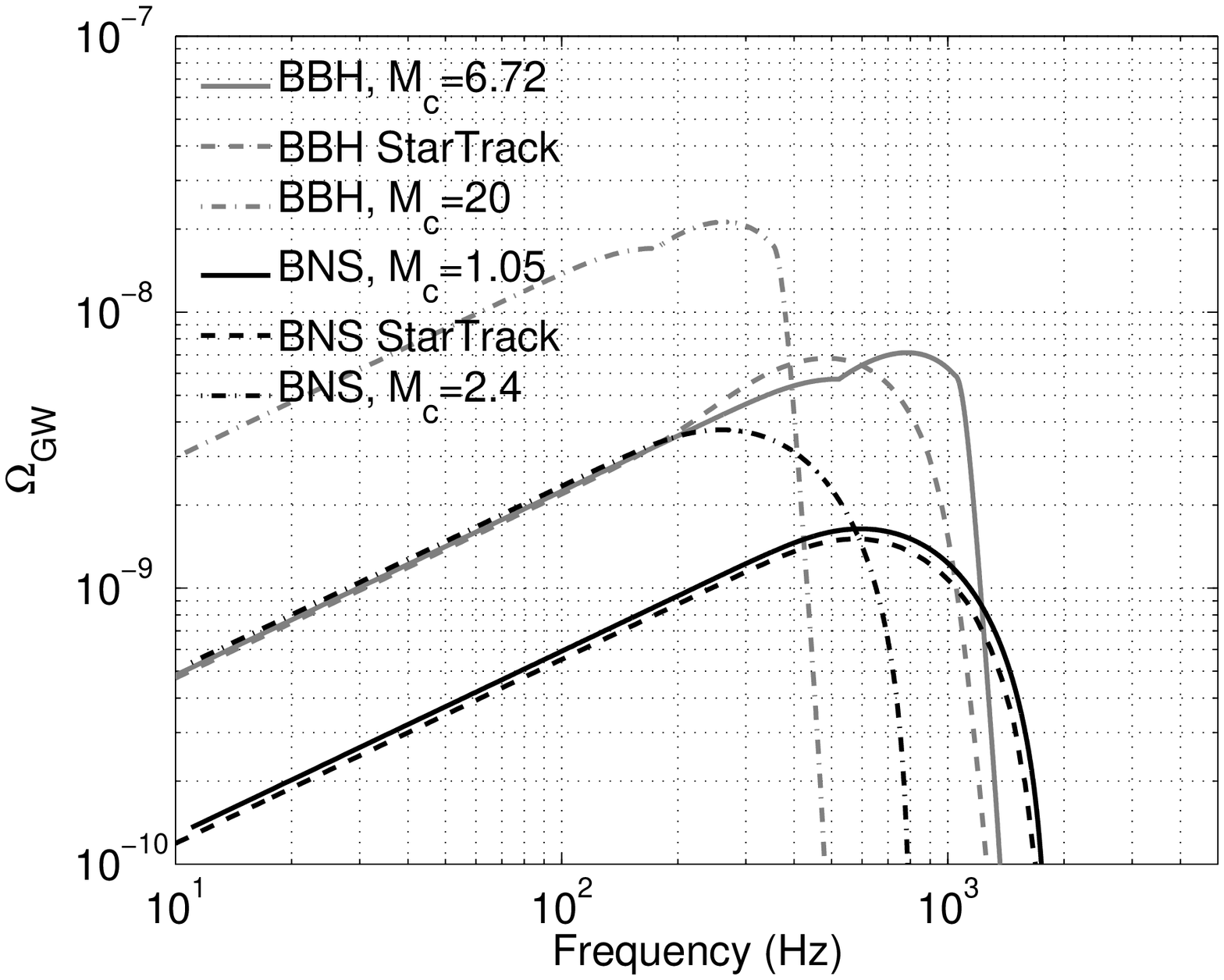, height=2.5in} &
    \psfig{file=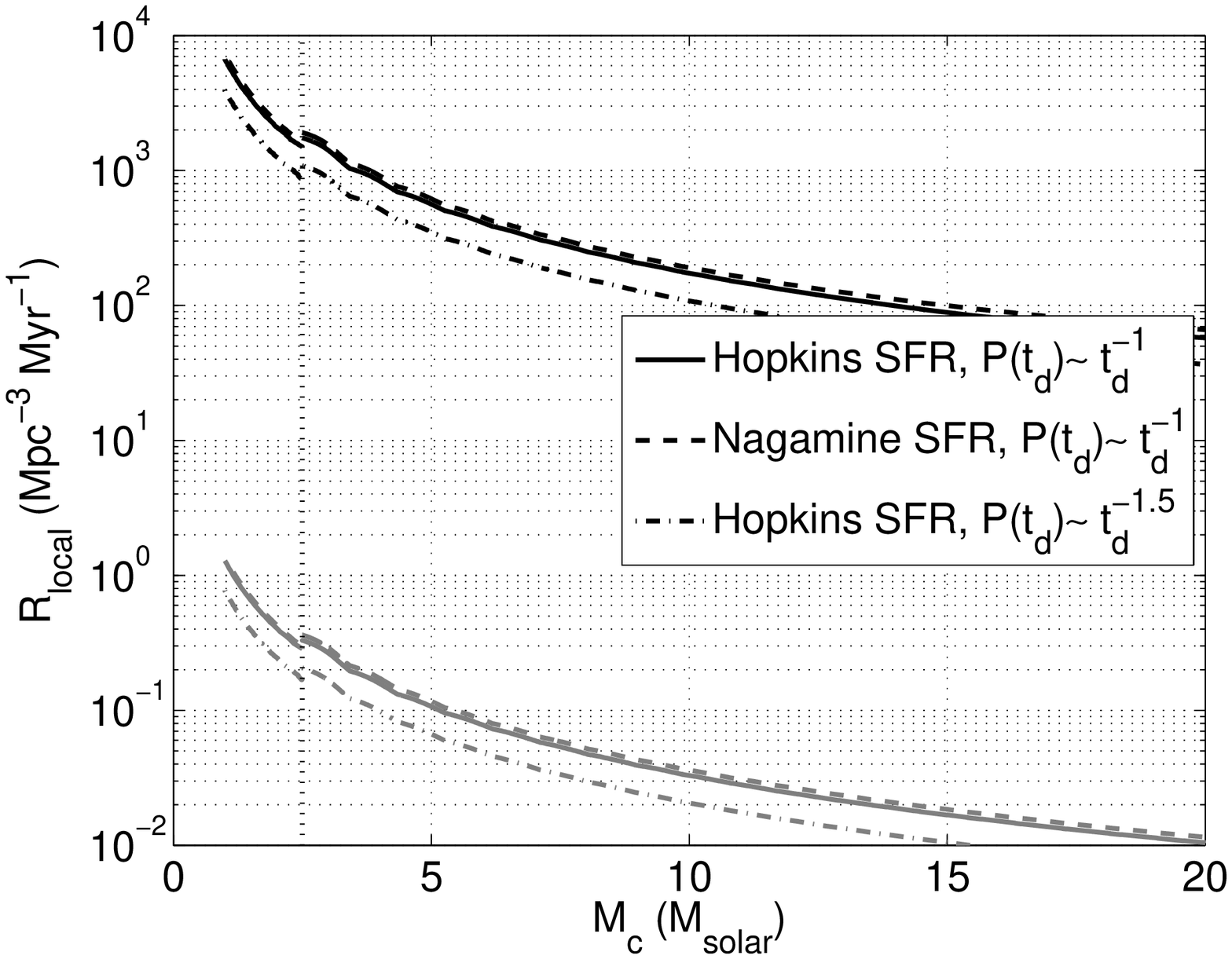, height=2.5in} \\
    \psfig{file=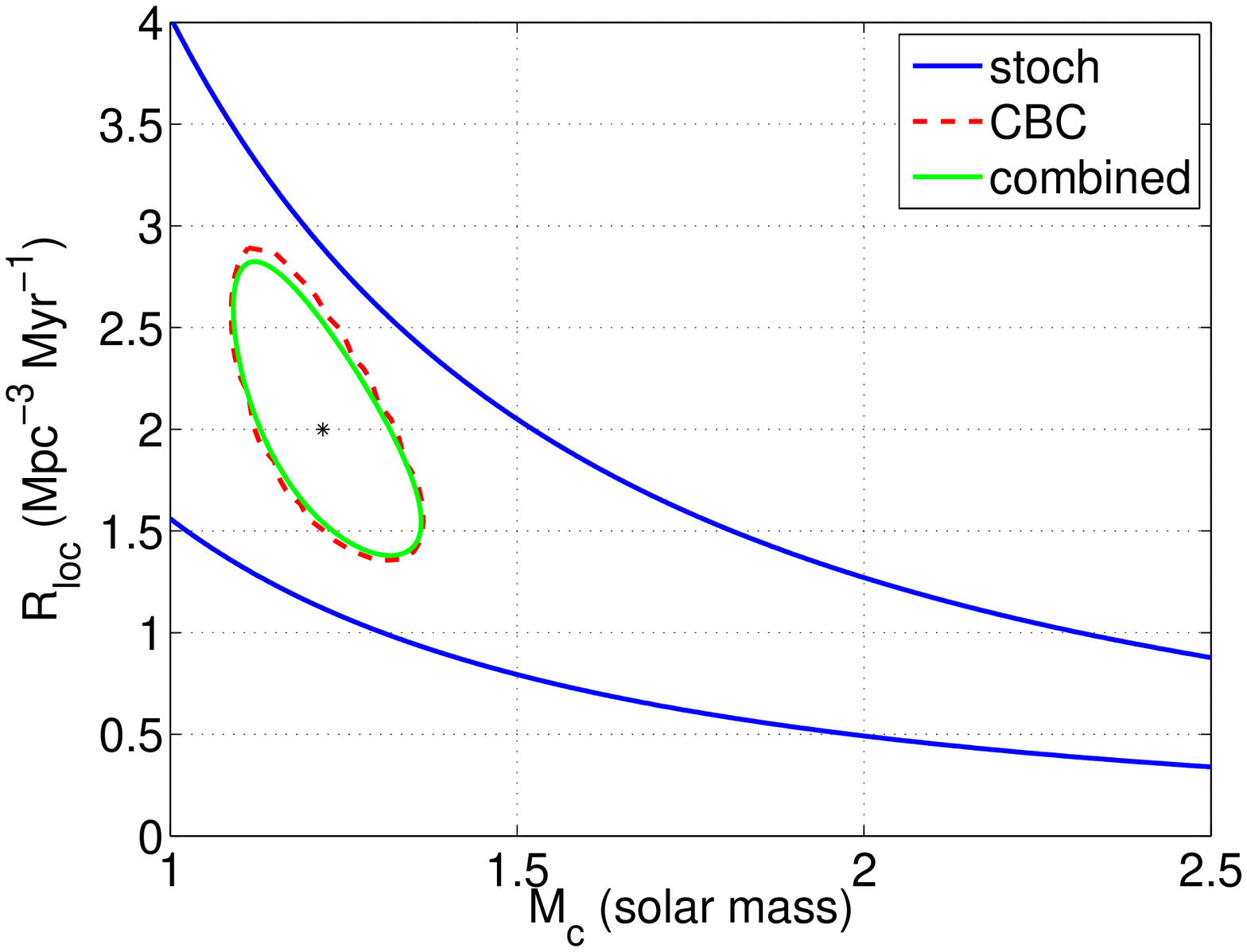, height=2.5in} &
    \psfig{file=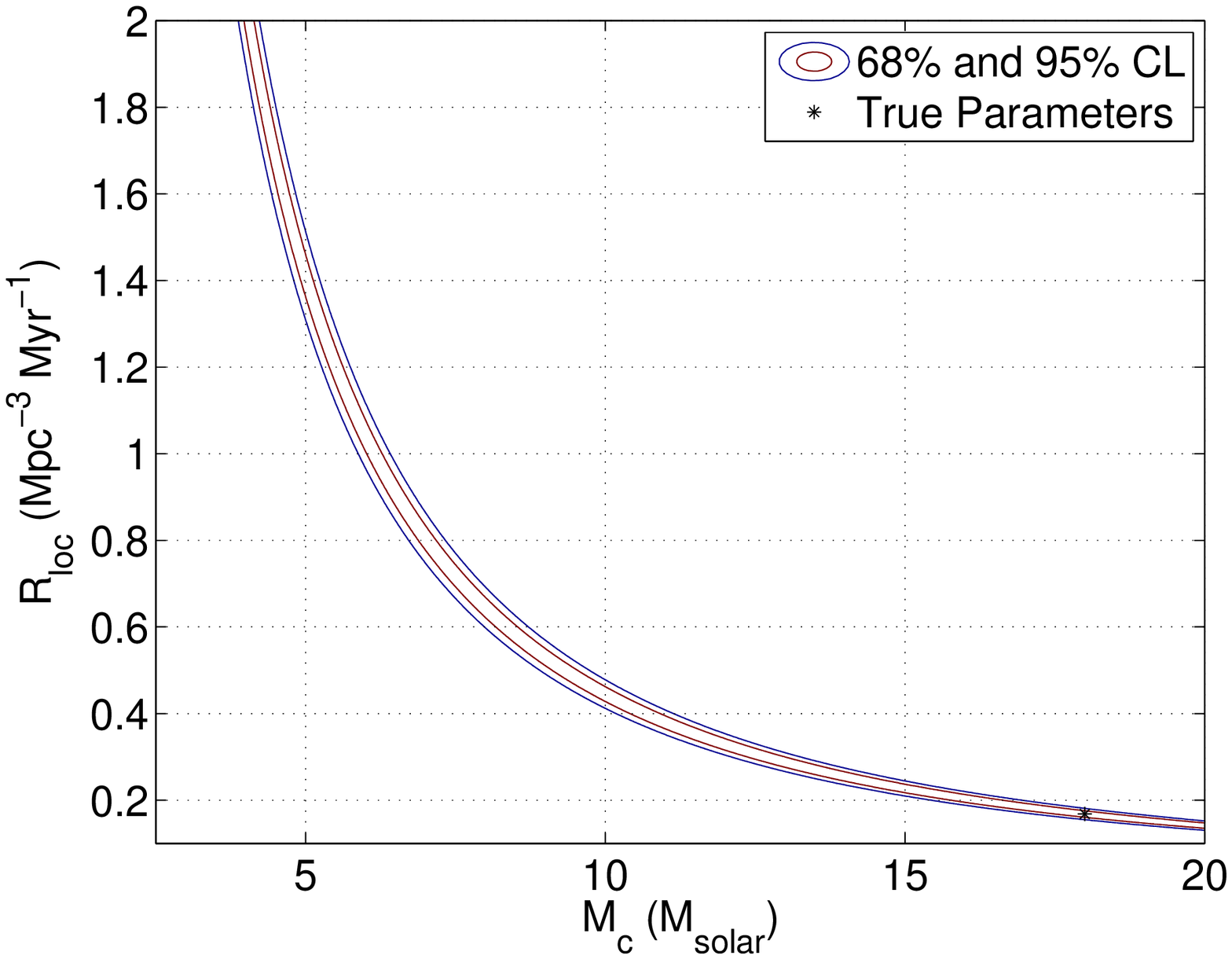, height=2.5in}
  \end{tabular}
   \caption{
     Top-left: GW spectra for several examples of CBC models are shown. The solid and dot-dashed curves were computed using the model described here, for the BNS (black) and BBH (gray) cases, while the dashed curves were computed using the distribution of chirp masses from StarTrack simulation \cite{startrack} with the same average chirp masses as those used for the solid curves. The solid and dashed curves are nearly identical for low values of $M_c$, indicating that using the average chirp mass is sufficient for the analysis presented here.
     Top-right: 95\% CL limits for CBC models for initial LIGO~\cite{S5stoch} and projected sensitivities for aLIGO assuming ${\rm SNR} = 2$ and $\unit[1]{yr}$ of coincident data from colocated interferometers operating at design sensitivity (and $\hat Y_i=0$). The region $M_c < 2.5 M_{\odot}$ corresponds to BNS models, and $M_c>2.5 M_{\odot}$ to BBH models. The three different styles (solid, dashed, and dot-dashed) correspond to three different choices of star formation rate $R_*(z)$ and time-delay probability density function $p(t_d)$, indicating the uncertainty due to these choices at the level of a factor of 2 \cite{StochCBC}.
     Bottom-left: projected 95\% CL contours for a simulated BNS signal measured with aLIGO.
     The simulated signal amplitude corresponds to local BNS rate in between the ``realistic'' and ``optimistic'' rates described in~\cite{cbc_rates}. Combining the stochastic measurement with direct observations of individual BNSs leads to a marginally tighter constraint on the astrophysical parameters describing the BNS population.
     Bottom-right: projected 95\% CL contours for a simulated stochastic BBH signal measured with aLIGO. A rather optimistic local rate $R_{\rm loc} = \unit[0.17]{Mpc^{-3}Myr^{-1}}$ was assumed for this simulation~\cite{cbc_rates}, resulting in a relatively narrow contours in the $R_{\rm loc}-M_c$ plane.
%
However, the signal is still not strong enough to differentiate between a high $R_\text{loc}$ - low $M_c$ signal and a low $R_\text{loc}$ - high $M_c$ signal.
%
   }
\label{fig:BNS}
\end{figure*}

Following \cite{StochCBC}, in the BNS case we include only the inspiral part of the GW signal, and write the GW spectrum as an integral over redshift $z$ and over the chirp mass $M_c'$:
\begin{widetext}
\begin{eqnarray}
\Omega_{\rm M}(f; M_c,\lambda) & = & \frac{8\lambda (\pi G)^{5/3}f^{2/3}}{9 H_0^3 c^2} \int dM_c' \; p(M_c') \; M_c'^{5/3}
 \int^{z_\text{sup}(M_c')}_{0} \frac{R_V(z) dz}{(1+z)^{1/3} E(\Omega_{\rm m},\Omega_{\Lambda},z)} \nonumber \\
& \approx & \frac{8\lambda (\pi G M_c)^{5/3}}{9 H_0^3 c^2}  f^{2/3} \int^{z_\text{sup}(M_c)}_{0} \frac{R_V(z) dz}{(1+z)^{1/3} E(\Omega_{\rm m},\Omega_{\Lambda},z)} .
\label{omega}
\end{eqnarray}
\end{widetext}
We have verified that the shape of the chirp mass distribution has a negligible effect on the GW spectrum, especially at frequencies below the peak where terrestrial GW detectors are most sensitive (see Fig. \ref{fig:BNS}). We therefore approximate the GW spectrum using only the average chirp mass $M_c$, as shown in the second line of Eq. \ref{omega}. The free parameters of the model are therefore $M_c$ and $\lambda$, the mass fraction parameter. We treat the BNS and BBH populations separately, with different average chirp masses. Furthermore, for the BBH case we use the more complex functional form derived by \cite{ajith} and used by \cite{zhu,StochCBC}, which includes the inspiral, merger, and ringdown contributions to the gravitational-wave signal (see \cite{zhu} for more detail).
The merger and ringdown spectra are computed assuming equal mass black holes.

The $\lambda$ parameter is proportional to the local CBC rate per unit volume (discussed further below), and it captures uncertainties associated with the mass fraction of the neutron star and black hole progenitors, the fraction of massive binaries formed among all stars, and the fraction of all binaries that remain bounded after the second supernova event. Further, $G$ is Newton's constant, $c$ is the speed of light, $E(\Omega_{\rm m},\Omega_{\Lambda},z) = \sqrt{\Omega_{\rm m} (1+z)^3 + \Omega_{\Lambda}}$ captures the dependence of the comoving volume on redshift (we use the standard $\Lambda$CDM cosmology, $\Omega_{\rm m} = 0.3$, $\Omega_{\Lambda} = 0.7$), and $R_V(z)$ is the observed rate of binary coalescences given by the following integral over the time-delay $t_d$ between creation and coalescence of the binary:
\begin{equation}
  R_V(z) = \int_{t_{\rm min}}^{t_{\rm max}} \frac{1}{1+z_f} \; R_*(t_c(z) - t_d) p(t_d) dt_d.
\end{equation}
Here $t_c(z)$ is the cosmic time to coalescence corresponding to redshift $z$, $z_f$ is the redshift at the formation time $t_c(z) - t_d$, and $R_*$ is the star formation rate (SFR). Multiple SFR models have been proposed \cite{hopkins,fardal,wilkins,nagamine,springel}. We adopt the model from~\cite{hopkins} and note that other SFR models can lead to variations in $\Omega_{\rm M}(f; M_c,\lambda)$ of up to a factor of two \cite{StochCBC}.
Finally, $p(t_d)$ is the probability density function for the time-delay---we use $p(t_d) \sim t_d^{-1}$, with the minimum time-delay $t_\text{min} = \unit[20]{Myr}$ for BNS and $\unit[100]{Myr}$ for BBH cases, and with $t_{\rm max}$ equal to the age of the universe.
We note that other choices of time-delay distribution could also lead to a factor of two difference \cite{StochCBC}. Finally, note the local CBC rate is given by $R_{\rm loc} = \lambda R_V(0)$.

The upper limit on the integral range in Eq. \ref{omega} depends on both the emission frequency range, $f_\text{min} - f_\text{max}$, in the source frame, and on the maximum redshift $z_\text{max}=6$ considered for the star formation history calculation:
\begin{equation}
  z_\text{sup} (f)= \left\lbrace
  \begin{array}{ll}
    z_\text{max}    &   \hbox{  if } f < f_\text{max} / (1+z_\text{max}) \\
    f_\text{max}/f - 1 &   \hbox{  otherwise }\\
  \end{array}
  \right.
  \label{eq-zsup}
\end{equation}

The top-left panel of Fig.~\ref{fig:BNS} shows several examples of BNS and BBH spectra---depending on the choice of parameter values, the spectra could peak in the sensitive band of aLIGO.
%
In the top-right panel of Fig.~\ref{fig:BNS} we show the 95\% CL limits in the $R_{\rm loc}-M_c$ plane (or, equivalently, $\lambda-M_c$ plane) obtained using recent LIGO results \cite{S5stoch} and the projected sensitivities for aLIGO at ${\rm SNR} = 2$, assuming $\unit[1]{yr}$ of exposure for colocated detectors operating at design sensitivity (and $\hat Y_i=0$)~\footnote{We note that, for these spectra, two colocated Advanced LIGO sensitivity interferometers are expected to have error bars on $\hat\Omega$ of roughly one half the size obtained from a network of interferometers consisting of KAGRA, LIGO H1, LIGO L1, VIRGO, and GEO.}.
Note that second-generation detectors are expected to be $10^4\times$ more sensitive to the CBC SGWB models than initial LIGO \cite{S5stoch}. Also note that, as stated above, different choices of the star formation rate and time-delay distribution lead to a factor of 2 uncertainty in the computed upper limits and projected sensitivities.

The bottom row of Fig.~\ref{fig:BNS} shows projected confidence contours for plausible simulated BNS (left) and BBH (right) signals that are within the reach of advanced detectors. For the BBH case, we consider the local BBH rate $R_{\rm loc} = \unit[0.17]{Mpc^{-3} Myr^{-1}}$, which is in between the realistic rate ($\unit[0.005]{Mpc^{-3} Myr^{-1}}$) and the optimistic rate ($\unit[0.3]{Mpc^{-3} Myr^{-1}}$) from \cite{cbc_rates}. This results in relatively narrow contours in the $R_{\rm loc}-M_c$ plane.
It is evident, however, that this signal is not strong enough to break the degeneracy between the $R_{\rm loc}$ and $M_c$ parameters---roughly 10 times higher rate of BBH systems is needed in order to break this degeneracy.
For the BNS case, we consider the local BNS rate $R_{\rm loc} = \unit[2]{Mpc^{-3} Myr^{-1}}$, which is in between the realistic rate ($\unit[1]{Mpc^{-3} Myr^{-1}}$) and the optimistic rate ($\unit[10]{Mpc^{-3} Myr^{-1}}$) from \cite{cbc_rates}. Again, the stochastic measurement alone is not sufficient to break the degeneracy between $R_{\rm loc}$ and $M_c$. However, we demonstrate in this case that our likelihood formalism can be naturally extended to include measurements of individually resolvable binary coalescences to extract more information about these parameters and about the source of the GW background.

The $R_{\rm loc}$-$M_c$ likelihood function from individual CBC detections is given approximately by
\begin{equation}\label{eq:CBC}
  \begin{split}
  L(\hat{n}, \hat\mu | R_{\rm loc}, M_c) \propto
  \exp\left[-\frac{(\hat\mu-M_c)^2}{2\sigma_\mu^2}\right] \\
  \frac{(V(M_c)R_{\rm loc} T)^{\hat{n}}}{\hat{n}!}
  e^{-V(M_c)R_{\rm loc} T} ,
  \end{split}
\end{equation}
where $\hat\mu$ is the average reconstructed chirp mass (with uncertainty $\sigma_\mu$), $\hat{n}$ is the number of CBC detections above threshold, $V(M_c)$ is the average volume of space in which CBC signals with chirp mass $M_c$ can be detected, and $T$ is the observation time.
Eq.~\ref{eq:CBC} does not take into account the fact that the detectability of CBC signals varies with chirp mass---SNR grows like $M_c^{5/6}$ \cite{finn}---but this introduces only a small bias $\lesssim4\%$.
The uncertainty $\sigma_\mu$ depends on the variance of the true chirp mass distribution, the number of measurements, and the measurement errors
associated with individual $M_c$ measurements, which are typically $\lesssim 0.5 M_\odot$ depending on the SNR.

We assume that the true chirp mass distribution is a delta function at $\mu=1.22 M_\odot$ (each NS has a mass of $1.4 M_\odot$). Given our ``semi-optimistic'' rate of $\unit[2]{Mpc^{-3} Myr^{-1}}$, we expect $80$ BNS events in 1 year of aLIGO data~\cite{cbc_rates}.
We further assume that each chirp mass can be reconstructed with a precision of $\approx0.5 M_\odot$ so that the uncertainty for the average is $\sigma_\mu = 0.5 M_\odot/\sqrt{80}\approx 0.06 M_\odot$.
(If the true chirp mass distribution has a width comparable to or greater than the detector resolution then $\sigma_\mu$ will grow appreciably.)

In the bottom-left panel of Fig.~\ref{fig:BNS} we include 95\% confidence contours obtained using just stochastic data, just direct BNS observations, and the contour obtained by computing the joint likelihood function as the product of likelihood functions given in Eqs. \ref{sgwblik} and \ref{eq:CBC}.
The joint likelihood leads to the tightest constraints on the model parameters, although in this particular example, the contour is determined primarily by the direct BNS observations.

Perhaps an even more important application of the joint likelihood technique will be in understanding the energy budget of the observed SGWB.
In our example, the agreement between the stochastic and CBC contours indicates that the observed SGWB is consistent with a single SGWB source, namely the BNS coalescences. In general, however, multiple SGWB sources (BNS, BBH, magnetars, and others) are expected to contribute to the observed SGWB.
%
By constructing a ``grand likelihood function'' with stochastic and transient measurements---and allowing for multiple SGWB models---it will be possible to identify contributions from different SGWB sources.
Understanding the energy budget of the observed SGWB may be critical for detection of the cosmological contribution, which will likely be masked by astrophysical SGWB sources in the frequency band of terrestrial GW detectors. Successful implementation of this technique will require simultaneous estimation of more than two parameters, which in turn will require more sophisticated techniques for likelihood maximization, such as Markov-chain Monte Carlo~\cite{mcmc} or nested sampling~\cite{multinest}. These approaches are currently being investigated.

{\em Conclusions.}---We have presented the first multi-dimensional confidence intervals for models of the stochastic gravitational-wave background.
The two cases we considered, generic power-law and compact binary coalescence models, can both be parametrized with just two parameters.
Other models, invoking, for example, magnetars or cosmic strings, are likely to require more than two parameters.
As the dimensionality of the likelihood function increases, techniques such as Markov-chain Monte Carlo~\cite{mcmc} or nested sampling~\cite{multinest} will likely be necessary in order to efficiently calculate confidence intervals on model parameters.

We have demonstrated that our technique can be extended to combine stochastic measurements with direct detections of gravitational-wave transients, yielding better constraints on astrophysical parameters. Furthermore, the technique provides a formalism for understanding the energy budget of the observed stochastic gravitational-wave background, identifying contributions due to different known astrophysical and cosmological sources and potentially revealing unexpected components.
This work (LIGO-P1200060) was supported by NSF grants
PHY-0758035 and
PHY-0970074.

\bibliography{paramest}

\end{document}